\documentclass[onecolumn]{aastex631}

\usepackage{comment}

\pdfoutput=1

\newcommand{\dgg}{$^{\circ}$}
\newcommand{\dg}{$^{\circ}$ }
\newcommand{\kms}{  {km~s$^{-1}$}}

\received{May 17, 2024}
\revised{July 15, 2024}
\accepted{July 16, 2024}

\shorttitle{Flux rope modeling of a CME observed by PSP and Solar Orbiter}
\shortauthors{E.E.Davies et al.}
\graphicspath{{./}{figures/}}

\begin{document}

\title{Flux rope modeling of the 2022 Sep 5 CME observed by Parker Solar Probe and Solar Orbiter from 0.07 to 0.69~au}

\author[0000-0001-9992-8471]{Emma E. Davies}
\affiliation{Austrian Space Weather Office, GeoSphere Austria, Graz, Austria}
\correspondingauthor{Emma E. Davies}
\email{emma.davies@geosphere.at}

\author[0000-0002-2559-2669]{Hannah T. Rüdisser}
\affiliation{Austrian Space Weather Office, GeoSphere Austria, Graz, Austria}
\affiliation{Institute of Physics, University of Graz, Graz, Austria}

\author[0000-0003-1516-5441]{Ute V. Amerstorfer}
\affiliation{Austrian Space Weather Office, GeoSphere Austria, Graz, Austria}

\author[0000-0001-6868-4152]{Christian Möstl}
\affiliation{Austrian Space Weather Office, GeoSphere Austria, Graz, Austria}

\author[0000-0002-2507-7616]{Maike Bauer}
\affiliation{Austrian Space Weather Office, GeoSphere Austria, Graz, Austria}
\affiliation{Institute of Physics, University of Graz, Graz, Austria}

\author[0009-0004-8761-3789]{Eva Weiler}
\affiliation{Austrian Space Weather Office, GeoSphere Austria, Graz, Austria}
\affiliation{Institute of Physics, University of Graz, Graz, Austria}

\author[0000-0001-9024-6706]{Tanja Amerstorfer}
\affiliation{Austrian Space Weather Office, GeoSphere Austria, Graz, Austria}

\author[0000-0002-6553-3807]{Satabdwa Majumdar}
\affiliation{Austrian Space Weather Office, GeoSphere Austria, Graz, Austria}

\author[0000-0003-1377-6353]{Phillip Hess}
\affiliation{U.S. Naval Research Laboratory, Washington, DC, USA}

\author[0000-0002-6273-4320]{Andreas J. Weiss}
\affiliation{NASA Postdoctoral Program Fellow, NASA Goddard Space Flight Center, Greenbelt, MD, USA}

\author[0000-0002-6362-5054]{Martin A. Reiss}
\affiliation{Community Coordinated Modeling Center, NASA Goddard Space Flight Center, Greenbelt, MD 20771, USA}

\author[0000-0002-0053-4876]{Lucie~M.~Green}
\affiliation{University College London, Mullard Space Science Laboratory, Holmbury St. Mary, Dorking, Surrey, RH5 6NT, UK}

\author[0000-0003-3137-0277]{David M.~Long}
\affiliation{School of Physical Sciences, Dublin City University, Glasnevin Campus, Dublin, D09 V209, Ireland}

\author[0000-0003-0565-4890]{Teresa Nieves-Chinchilla}
\affiliation{Heliophysics Science Division, NASA Goddard Space Flight Center, Greenbelt, MD, USA}
\affiliation{Department of Physics and Astronomy, George Mason University, Fairfax, VA, USA}

\author[0000-0002-0608-8897]{Domenico Trotta}
\affiliation{Imperial College London, South Kensington Campus, London SW7 2AZ, UK}

\author[0000-0002-7572-4690]{Timothy S. Horbury}
\affiliation{Imperial College London, South Kensington Campus, London SW7 2AZ, UK}

\author[0000-0002-9833-4097]{Helen O'Brien}
\affiliation{Imperial College London, South Kensington Campus, London SW7 2AZ, UK}

\author[0000-0002-4200-892X]{Edward Fauchon-Jones}
\affiliation{Imperial College London, South Kensington Campus, London SW7 2AZ, UK}

\author[0000-0001-5127-9273]{Jean Morris}
\affiliation{Imperial College London, South Kensington Campus, London SW7 2AZ, UK}

\author[0000-0002-5982-4667]{Christopher J. Owen}
\affiliation{University College London, Mullard Space Science Laboratory, Holmbury St. Mary, Dorking, Surrey, RH5 6NT, UK}

\author[0000-0002-1989-3596]{Stuart D. Bale}
\affiliation{Physics Department and Space Sciences Laboratory, University of California, Berkeley, USA}

\author[0000-0002-7077-930X]{Justin C. Kasper}
\affiliation{School of Climate and Space Sciences and Engineering, University of Michigan, Ann Arbor, Michigan, USA}

\begin{abstract}

As both Parker Solar Probe (PSP) and Solar Orbiter (SolO) reach heliocentric distances closer to the Sun, they present an exciting opportunity to study the structure of CMEs in the inner heliosphere. We present an analysis of the global flux rope structure of the 2022 September 5 CME event that impacted PSP at a heliocentric distance of only 0.07~au and SolO at 0.69~au. We compare in situ measurements at PSP and SolO to determine global and local expansion measures, finding a good agreement between magnetic field relationships with heliocentric distance, but significant differences with respect to flux rope size. We use PSP/WISPR images as input to the ELEvoHI model, providing a direct link between remote and in situ observations; we find a large discrepancy between the resulting modeled arrival times, suggesting that the underlying model assumptions may not be suitable when using data obtained close to the Sun, where the drag regime is markedly different in comparison to larger heliocentric distances. Finally, we fit the SolO/MAG and PSP/FIELDS data independently with the 3DCORE model and find that many parameters are consistent between spacecraft, however, challenges are apparent when reconstructing a global 3D structure that aligns with arrival times at PSP and Solar Orbiter, likely due to the large radial and longitudinal separations between spacecraft. From our model results, it is clear the solar wind background speed and drag regime strongly affect the modeled expansion and propagation of CMEs and need to be taken into consideration.

\end{abstract}


\section{Introduction} \label{sec:intro}

Coronal mass ejections (CMEs) are large-scale expulsions of plasma and magnetic fields that are driven from the solar atmosphere. As they propagate and expand outwards through the heliosphere, the heliospheric counterparts to CMEs measured in situ by spacecraft, are known as interplanetary coronal mass ejections (ICMEs). ICMEs can be distinguished from the ambient solar wind by features such as an enhanced magnetic field strength, low plasma $\beta$, declining velocity profile and decreased proton temperature amongst many others \citep[e.g.][]{zurbuchen2006situ}. 

Until recently, in situ properties of ICMEs have been studied over varying heliocentric distances, mostly between 0.3 and 5.4~au, utilising both dedicated solar wind missions e.g. Helios \citep[0.3--1~au, e.g.][]{cane1997helios, bothmer1998structure}, ACE/Wind, STEREO \citep[at 1~au, e.g.][]{cane2003interplanetary, richardson2010near, jian2018stereo} and Ulysses \citep[1--5.4~au, e.g.][]{liu2005statistical, wang2005characteristics, ebert2009bulk, du2010interplanetary, richardson2014identification}, as well as planetary missions when outside of their respective planetary environments such as BepiColombo \citep{davies2021solo, moestl2022multi, palmerio2024bepi}, MESSENGER \citep{winslow2015interplanetary, good2016interplanetary}, Venus Express \citep{good2016interplanetary}, and Juno \citep[1--5.4~au, e.g.][]{davies2021catalogue}. Unfortunately, many of the planetary mission measurements of ICMEs only consist of magnetic field observations, often lacking complementary measurements of the solar wind plasma. However, with the addition of the Solar Orbiter \citep[SolO;][]{mueller2020solar} and Parker Solar Probe \citep[PSP;][]{fox2016psp} missions, we are now able to study ICMEs much closer to the Sun, with a full-suite of instruments capable of analyzing their properties and structure in detail.

The generally accepted picture of the global structure of an ICME comprises a magnetic flux rope, where helical field lines are wound around a central axis, which is driven outwards from the Sun where both legs remain connected to the solar corona. If traveling faster than the local magnetosonic speed, such a structure drives a forward shock, behind which there is a turbulent and compressed region of magnetic field known as the sheath \citep[see, e.g., Figure 2 in][]{zurbuchen2006situ}. Such ICMEs, with a clear magnetic flux rope structure, are often described as magnetic clouds, where an enhanced magnetic field, smooth rotation of the magnetic field vector, low plasma $\beta$ and a drop in proton temperature \citep{burlaga1981magnetic} are observed in situ. However, not all ICMEs are observed to have such a magnetic cloud structure; \citet{gosling1990coronal} classified $\sim$ 30\% of ICMEs at 1~au as magnetic clouds, the proportion varying with the solar cycle \citep{cane2003interplanetary}. For events where all the features necessary to be classified as a magnetic cloud are not met, studies often use the more general term of magnetic ejecta (ME) to describe this interval.

As in situ measurements of ICMEs are often taken by just a single spacecraft, observations of the rotation of the magnetic field taken along the single 1D trajectory can be interpreted as the spacecraft measuring the magnetic flux rope \citep{burlaga1981magnetic, marubashi1986structure,burlaga1988magnetic}, and the 3D global structure recreated by fitting the magnetic field vector with force-free flux rope models such as the Lundquist \citep{lundquist1950} or Gold-Hoyle solutions \citep{gold1960origin}. 

However, in reality, reconstructing the global structure of ICMEs is more complex; as ICMEs propagate through the heliosphere, they may undergo many processes including rotations, deflections, interactions with other solar wind transients and erosion via magnetic reconnection \citep[see review by][]{manchester2017physical}. A study by \citet{scolini2022complexity} found that fundamental alterations or re-orientations of the ICME occurred for $\sim$65\% of ICMEs as they propagated between 0.3 and 1~au, attributing the complication in their evolution primarily to interactions with other large-scale solar wind structures. Even simply propagating through the ambient and structured solar wind, deformations of the ICME cross-section can occur \citep{riley2004kinematic, owens2006kinematically, savani2010observational, davies2021solo}. There are also many questions that remain about the global picture of ICMEs affecting their reconstruction with flux rope models e.g.: to what extent can ICMEs be considered coherent structures \citep[e.g.][]{owens2017coronal, scolini2022complexity}, and do the ICME legs also contain a twisted flux rope structure (e.g. \citealt{owens2016legs}, \citealt{ruedisser2024understanding} (ApJ, in revision))?

To truly understand the global structure of ICMEs and their evolution as they propagate through the heliosphere, multi-spacecraft measurements of the same ICME over varying radial and longitudinal separations are necessary \citep{lugaz2018}. Multi-spacecraft ICME studies such as \citealt{good2018correlation}, \citealt{salman2020radial}, \citealt{davies2022multi}, and \citealt{moestl2022multi} took advantage of conjunctions between spacecraft including MESSENGER, Venus Express, Wind, ACE, STEREO-A, Juno, SolO, BepiColombo, and Parker Solar Probe, where an ICME was observed by more than one spacecraft. Measurements of the same ICME taken during these different spacecraft configurations mean that ICME properties are often tracked over a varying range of radial and longitudinal separations, making it hard to disentangle radial and longitudinal (and in some cases, latitudinal) effects on ICME evolution. These studies found that although the evolution of ICME properties with increasing heliocentric distance was consistent with previous statistical trends, individual ICMEs display significant variability when compared to average trends.

The CME launched on 2022 September 5 is the subject of great interest \citep[e.g.][]{long2023eruption, romeo2023nearsun, patel2023wispr, trotta2024properties, liu2024direct}, as it is the closest CME observed to Sun to date, observed by PSP at a heliocentric distance of just 0.07~au, and later by SolO at 0.69~au. Such a spacecraft configuration presents a unique opportunity for a direct comparison of ICME observations at an early stage of its evolution, with those much further in its evolution. \citet{long2023eruption} focused on the solar source of the CME, identifying plasma flows along a filament channel to infer the magnetic field configuration of the flux rope and relate these to the in situ observations at PSP. The location of PSP so close to the Sun provides a direct intersection between remote and in situ observations;  \citet{romeo2023nearsun} compared the in situ observations at PSP to remote-sensing observations, and found that due to the complexity of the ICME, it was hard to reconcile the observations into one simple description of the event. Focusing on the in situ PSP observations, \citet{romeo2023nearsun} and \citet{liu2024direct} also found that the CME played a significant role in the evolution of the heliospheric current sheet (HCS).

In this study, we compare the observations at PSP with those at SolO. \citet{trotta2024properties} previously compared observations of the ICME shock, comparing typical shock parameters and small-scale features at both spacecraft, revealing important differences in the local environment in which the shocks were propagating. Here, we aim to reconcile observations at PSP with those at SolO to reconstruct the global structure of the ICME. In Section \ref{sec:data}, we present the in situ magnetic field and solar wind plasma data at each spacecraft. In Section \ref{sec:ELEvoHI}, we use remote images taken by PSP to model the arrival time at the two spacecraft using the ELEvoHI model, and Section \ref{sec:3DCORE} presents the results of fitting the magnetic field data at both spacecraft with the 3DCORE model to reconstruct the global structure of the ICME flux rope. 

\section{Spacecraft Data and Observations} \label{sec:data}

\subsection{Spacecraft data}

To identify the ICME in situ, we present magnetic field \citep[FIELDS;][]{bale2016fields} observations and solar wind plasma \citep[SWEAP;][]{kasper2016sweap} data measured by the Solar Probe ANalyzer for Ions \citep[SPAN-I;][]{livi2022spani} instrument at Parker Solar Probe \citep[PSP;][]{fox2016psp}, as well as measurements of the magnetic field \citep[MAG;][]{horbury2020mag} and solar wind plasma \citep[SWA;][]{owen2020swa} at Solar Orbiter \citep[SolO;][]{mueller2020solar}. We are particularly thankful to the Solar Orbiter MAG team, who have provided updated files replacing those with previously missing data during the event. These files are available on request from the MAG team and will be publicly available on the ESA Solar Orbiter Archive soon (\url{https://soar.esac.esa.int/soar/}).

In addition to the in-situ data, this work also incorporates the use of data from various remote sensing instruments, presenting the solar source of the in situ observations. Disk imaging observations from the 17.4~nm passband of the Full Sun Imager (FSI), one of the telescopes of the Extreme Ultraviolet Imager \citep[EUI;][]{rochus2020eui} suite onboard SolO, are shown. Coronagraph images from the Large Angle Spectroscopic COronagraph C2 \citep[LASCO;][]{brueckner1995lasco} onboard the Solar and Heliospheric Observatory \citep[SOHO;][]{domingo1995soho}, and the outer coronagraph COR-2 onboard the Solar Terrestrial Relations Observatory \citep[STEREO;][]{kaiser2008stereo} are presented. Images taken by the Wide-Field Imager for Solar Probe \citep[WISPR;][]{vourlidas2016wispr} onboard PSP are also used to model CME arrival times in Section \ref{sec:ELEvoHI}.

\subsection{Spacecraft configuration}

\begin{figure}
\centering
{\includegraphics[width=\textwidth]{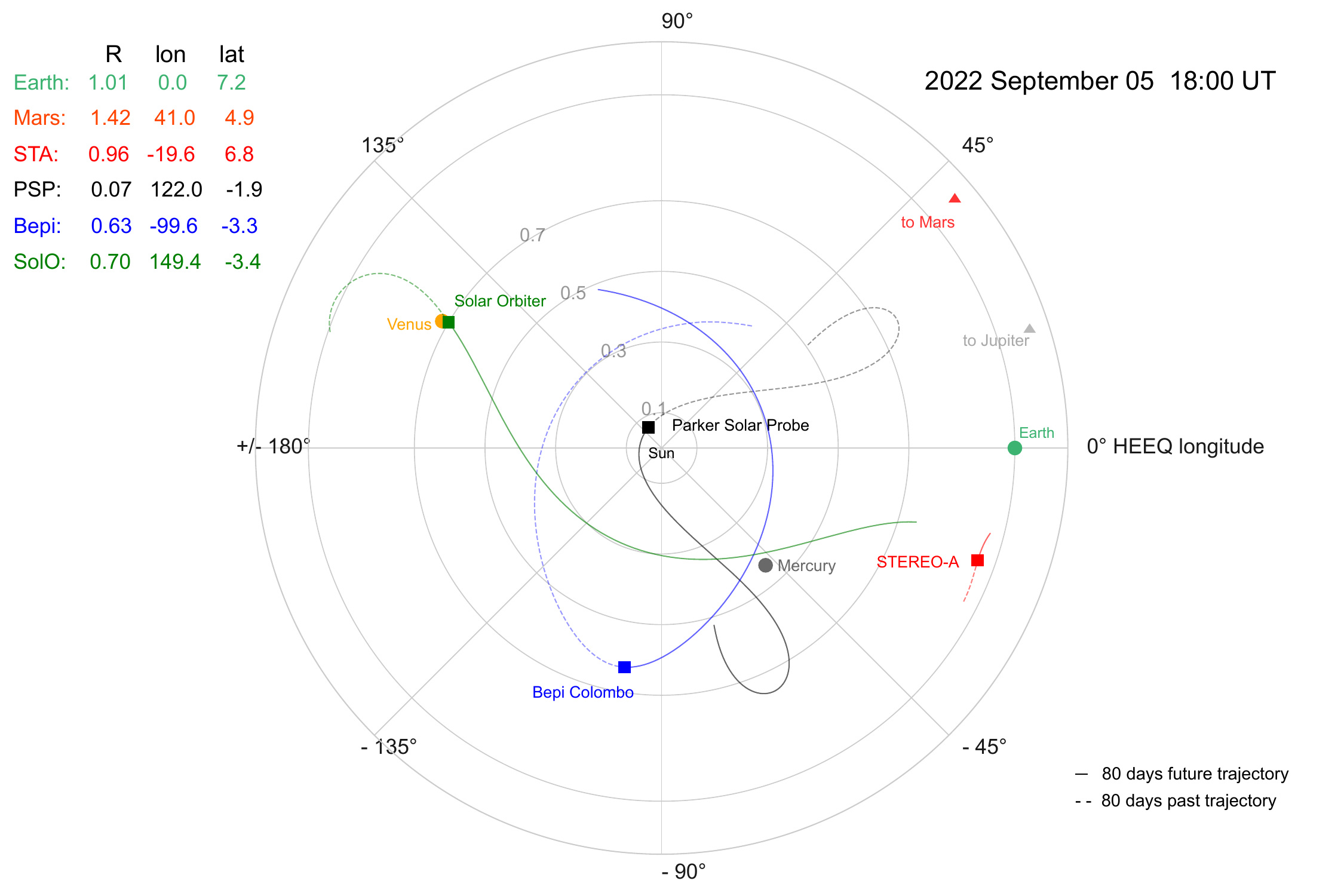}}
\caption{Spacecraft positions presented in Heliocentric Earth Equatorial (HEEQ) coordinates, on 2022 September 5 18:00 UT, around the time of arrival of the CME at PSP. SolO is represented by the green square marker, and PSP by the black square marker. The spacecraft trajectories shown cover a time range of 80 days before (dashed) and after (solid) the event. Exact spacecraft positions for defined ICME boundaries during the event are given in Table \ref{tab:insitu_params}.} 
\label{fig:sc_positions}
\end{figure}

Figure \ref{fig:sc_positions} presents an overview of the spacecraft configuration within 1~au on 2022 September 5 18:00~UT, around the time the ICME was observed at PSP. At this time, PSP (represented by the black square marker) was located at a heliocentric distance of 0.07~au and longitude of 122.1$^{\circ}$ in Heliocentric Earth Equatorial (HEEQ) coordinates, and SolO (represented by the green square marker) was located at a heliocentric distance of 0.70~au and longitude of 149.2$^{\circ}$. PSP and SolO were thus well aligned to observe the same ICME, with a longitudinal separation within 30$^{\circ}$ throughout the event. 

Such a configuration occurred due to SolO performing a gravity assist maneuver (GAM) at Venus on September 4, whilst PSP was undergoing Encounter 13, reaching a perihelion of 0.062~au on September~6. Figure \ref{fig:sc_positions} shows that both spacecraft were located on the far side of the Sun with respect to Earth, with STEREO-A located at -19.5$^{\circ}$ and BepiColombo at -99.6$^{\circ}$ HEEQ longitude. Compared to the positions of PSP and SolO, these separations are wider than typical longitudinal extents of ICMEs \citep[e.g.][]{good2016interplanetary, lugaz2024width}. We checked the magnetic field data at both spacecraft, and while STEREO-A did not observe any ICME signature, BepiColombo saw a shock and sheath on early September 7. However, these signatures may also be caused by a different CME, but this observation could warrant further investigation in connection with the event discussed here.

\subsection{Solar source of the in situ observations} \label{subsec:solar_source}

\begin{figure*}
\gridline{\fig{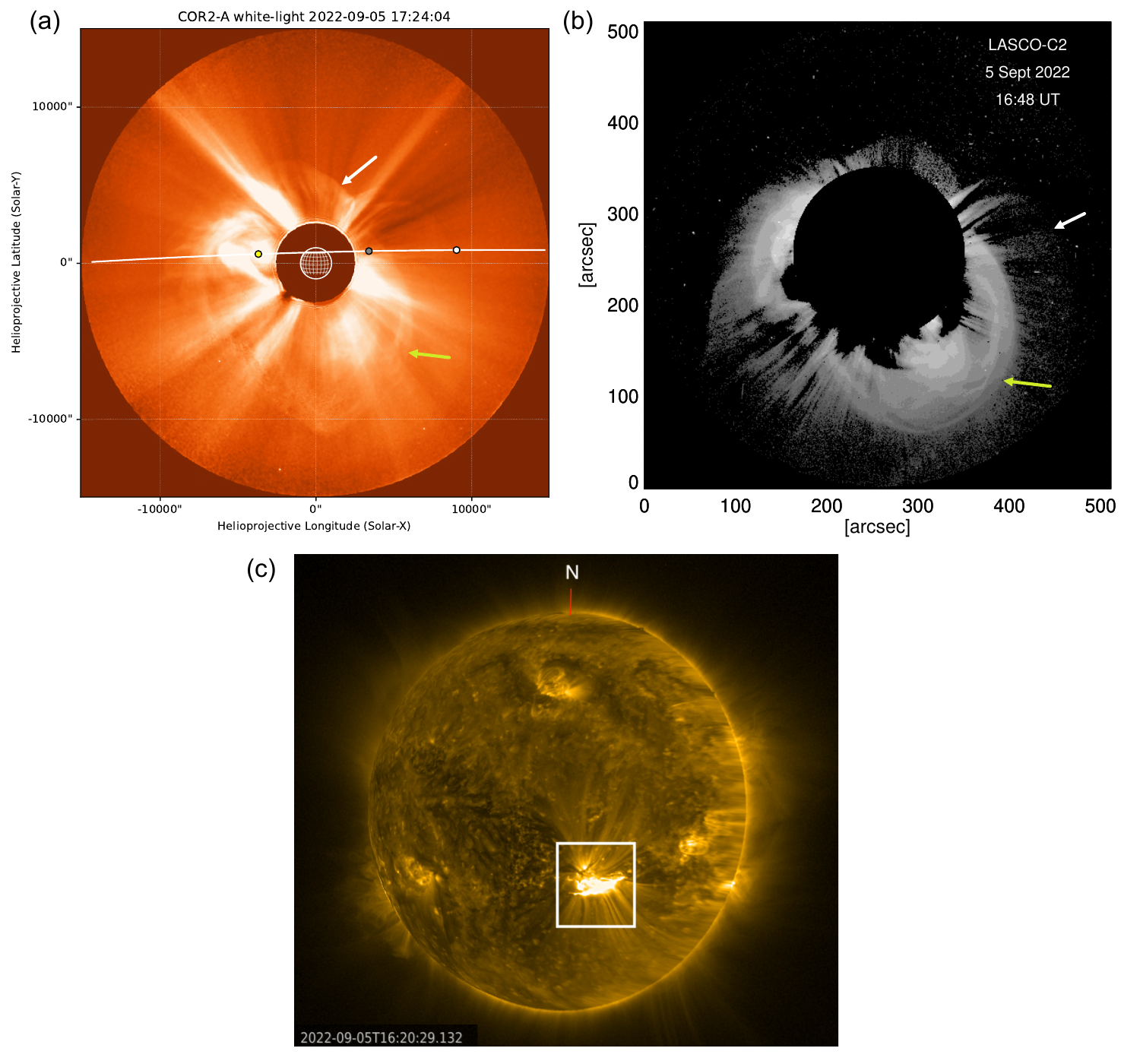}{0.95\textwidth}{}}
\caption{Coronagraph and solar observations. (a) The halo CME as observed by STEREO-A/COR-2 at 2022 Sep 5 17:24 UT, with the trajectory of PSP as the white solid line. The white dot indicates the PSP position at the arrival time of the CME shock at PSP (2022 Sep 5 17:27 UT), the grey dot the position at the PSP ME end time (2022 Sep 6 00:57), and the yellow dot the PSP position at the shock arrival time at SolO (2022 Sep 6 10:01). The yellow and white arrows point at the flux-rope and the white light shock signature associated with the CME. (b) The same CME as observed in the SOHO/LASCO-C2 field of view at 16:48 UT. The white and yellow arrows indicate the shock front and the flux-rope associated with the CME. (c) The source region of the CME (denoted inside the white rectangle) as observed by the first SolO/EUI telescope FSI in the 17.4~nm passband at 2022 Sep 5 16:20~UT.
\label{solar}}
\end{figure*}

On 2022 September 5, the CME was observed as a full halo event in the field of view of the coronagraph STEREO-A/COR-2. 
Figure~\ref{solar}a shows the coronagraph image for the CME at  17:24~UT in the COR-2 field of view and Figure~\ref{solar}b shows the CME as observed in the LASCO-C2 field of view at 16:48~UT. The embedded flux rope and the associated shock front can be identified (shown in yellow and white arrows respectively) in the two images. The white solid line in Figure~\ref{solar}a indicates the trajectory of PSP. The white and grey dots represent the arrival time of the CME associated shock and the end time of the magnetic ejecta (ME) at PSP respectively, while the yellow dot denotes the PSP position at the shock arrival time at SolO.

Coronagraph images are however, not always sufficient to clearly understand the propagation direction of a CME. This is particularly important for halo CMEs, as the determination of the propagation direction is crucial to identify if the CME is traveling towards or away from the Earth, and the halo appearance of the CME front makes it more tricky to ascertain the direction. In such cases, identifying the source region of the halo CME on the solar disk provides conclusive evidence on whether the CME was traveling towards or away from the Earth. The source region of the CME in this study was observed by the FSI telescope (in the 17.4~nm passband) of the EUI payload onboard SolO. The source region was identified by creating running difference images using the JHelioviewer software \citep{mueller2017jhelioviewer}. The erupting active region was located at a latitude of -25$^{\circ}$ and a longitude of 180$^{\circ}$ (denoted by a white box in Figure~\ref{solar}c) in Stonyhurst heliographic coordinates \citep[for details of the source region, please refer to][]{majumdar2023coronal}. The Stonyhurst coordinate system is directly relatable to the HEEQ system, and therefore, we expect the CME to propagate in a southwards direction with respect to the solar equatorial plane, and that PSP and SolO will likely observe the flank of the CME. 

The identified source region is in agreement with that identified by \citet{long2023eruption}. They studied the source region in detail, and identified a filament channel rooted in the strong magnetic field of the active region which stretched out across the disk into the quiet Sun. From their observations, they were able to infer the right-handed rotation of the magnetic field early in its evolution using the plasma flow observed along the channel. The active region itself also shows a right-handed nature, for example, through the flare loops that formed as a result of the reconnection associated with the CME \citep[see][and references therein for determining handedness of active regions]{palmerio2017determining}.

Although the CME was observed by both the LASCO-C2 and STEREO-A/COR-2 coronagraphs, only a couple of frames were available from LASCO due to a data outage, and therefore, STEREO-A/COR-2 was the only imager with continuous observation of the CME. Using the STEREO-A/COR-2 and source region information observed by SolO EUI/FSI, \citet{patel2023wispr} was able to constrain and implement the Graduated Cylindrical Shell model \citep[GCS;][]{thernesien2009forward} to the COR-2 images. The results of the GCS fitting indicated a southwards propagation direction, with a latitude of $\sim$~-18\dg and a longitude of $\sim$~166\dgg, with an axis tilt of $\sim$~58\dg (R. Patel, private correspondence).

\subsection{In situ spacecraft observations at PSP and Solar Orbiter} 
\label{subsec:insitu_obs}

\begin{table}
\centering
\begin{tabular}{l|cc}
\hline
 & \textbf{PSP} & \textbf{SolO}\\
\hline
ICME start, t$_1$ [UT] & 2022-09-05 17:27 & 2022-09-06 10:01 \\
ME start, t$_2$ [UT] & 2022-09-05 18:50 & 2022-09-07 01:25 \\
ME end, t$_3$ [UT] & 2022-09-06 00:57 & 2022-09-08 04:10 \\
\hline
R$_{1,2,3}$ [AU] & 0.070, 0.068, 0.063  & 0.695, 0.689, 0.677 \\
lat$_{1,2,3}$ [$^{\circ}$] & -1.83, -2.06, -3.05  & -3.55, -3.66, -3.86 \\
lon$_{1,2,3}$ [$^{\circ}$] & 120.72, 124.67, 144.20  & 149.66, 149.91, 150.38 \\
\hline
ME $\langle$B$\rangle$ [nT] & 896.9 & 15.8 \\
ME B$_{max}$ [nT] & 1167.9 & 33.8 \\
ME $\langle$V$\rangle$ [\kms] & 697.0 & 684.0 \\
ME $\langle$N$_p\rangle$ [cm$^{-3}$] & 1082.2 & 11.2 \\
ME $\langle$T$_p\rangle$ [MK] & 2.98 & 0.00 \\
\hline
\end{tabular}
\caption{ICME properties at PSP and Solar Orbiter. ICME shock, ME start and end times are given, along with the mean (maximum) magnetic field strength measured within the ME. Plasma properties i.e. bulk velocity, temperature and density, within the ME are given in the same format. The spacecraft positions in HEEQ (radial distance, latitude, and longitude) corresponding to the three ICME boundary times defined are also listed.}
\label{tab:insitu_params}
\end{table}

Table \ref{tab:insitu_params} summarizes the arrival times of the ICME (in this case, the shock arrival time, $t_1$), the start of the magnetic ejecta ($t_2$), and the end of the magnetic ejecta ($t_3$) at both PSP and SolO. The spacecraft heliocentric distance, latitude, and longitude (in HEEQ coordinates) at those times are also listed, as well as mean and maximum ICME parameters calculated over the duration of the magnetic ejecta at each spacecraft.

\begin{figure*}
\centering
{\includegraphics[width=\textwidth]{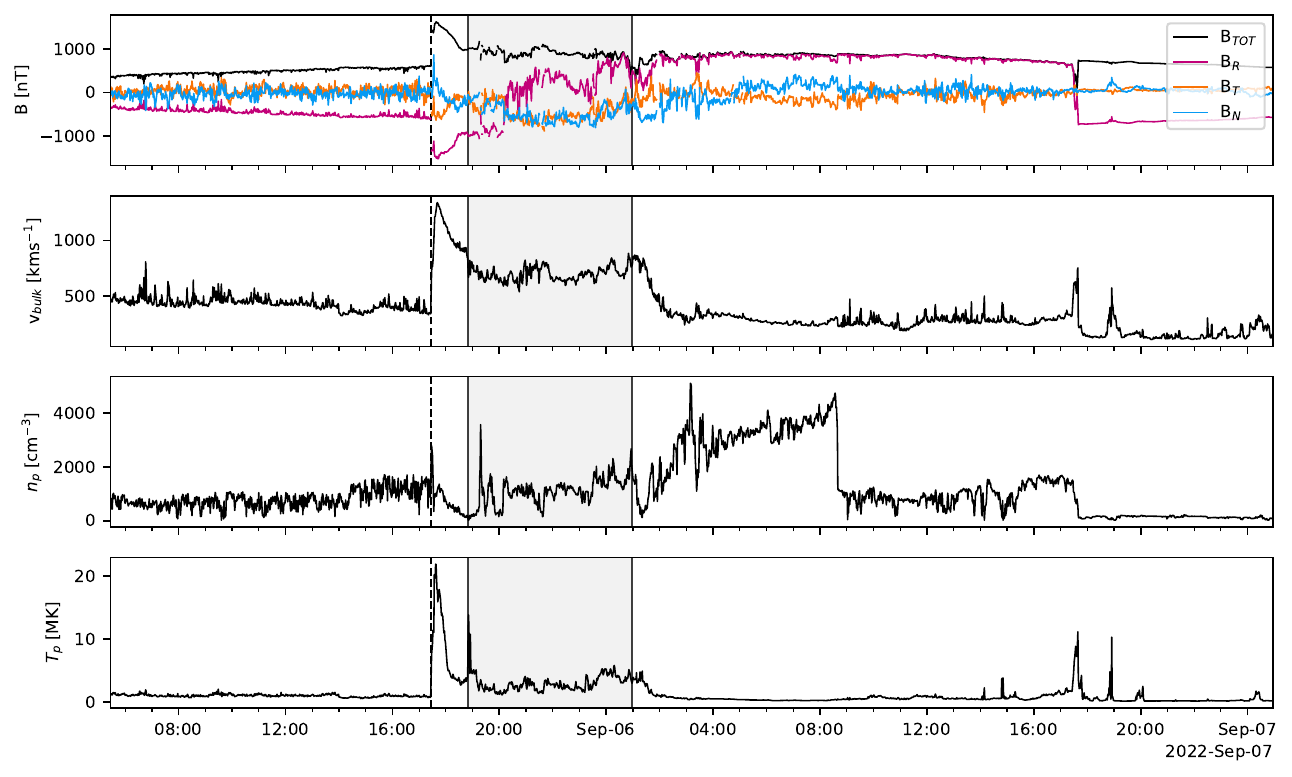}}
\caption{PSP in situ magnetic and plasma data. The top panel presents the magnetic field data, where the magnetic field magnitude is shown in black, and the magnetic field components in Radial-Tangential-Normal (RTN) coordinates in magenta, orange, and blue, respectively. The following panels present the bulk proton velocity, proton density, and proton temperature of the solar wind, respectively. The vertical dashed line mark the time of the shock front, and the solid vertical lines mark the start and end of the ME, which is also represented by the grey shaded area.} 
\label{fig:psp_insitu}
\end{figure*}

Figure \ref{fig:psp_insitu} presents the magnetic field and plasma measurements taken by PSP during the event. We observe a fast-forward shock driven by the CME at 17:27~UT, characterized by a sudden jump in magnetic field magnitude from $\sim$600~nT to 1600~nT, proton bulk velocity from 340 to 1340\kms, and proton temperature from 0.8 to 22~MK. The shock was observed by PSP while located at a heliocentric distance of 0.07~au ($\sim$15~$R_{\odot}$), making this the closest ICME to the Sun observed to date. Analysis conducted by \citet{trotta2024properties} used the Mixed Mode 3 method \citep[MX3;][]{paschmann2000multipoint} and a systematic variation of upstream/downstream averaging windows~\citep[see][]{Trotta2022btechniques} to determine an average shock normal of $n = (0.5, -0.8, 0.2)$ (in RTN coordinates), where $\theta = 53^\circ$ and therefore, the shock is oblique and consistent with a scenario in which PSP crosses the flank of the ICME \citep[see discussion and Figure 9a in][]{romeo2023nearsun}.

We define the start of the ME at 2022-09-05 18:50~UT, coincident with a sharp change in the bulk proton velocity from 850 to 670\kms, and the end of the ME at 2022-09-06 00:57~UT. This ME interval was defined as such to capture most of the rotation of the magnetic field observed by PSP, in this case, the radial magnetic field component (B$_R$), and is consistent with the boundaries defined using the $H_m$-PVI technique in \citet{long2023eruption}. Across the ME interval, the magnetic field magnitude remains fairly constant, with a mean magnetic field strength of 896.9~nT. The plasma properties i.e. the bulk proton velocity, proton density and proton temperature also all remain at consistent values during this interval (mean values of 697\kms, 1082~cm$^{-3}$, 2.98~MK, respectively), before returning to pre-ICME levels in the case of the bulk velocity and the temperature, and rising further in the case of the density. We attribute the sharp drop in density following this rise to the core of the ion distribution being largely outside of the SPAN-i field of view \citep{livi2022spani}. These plasma properties are not the typical features observed near 1~au described in \citet{zurbuchen2006situ} e.g. a declining speed profile, low proton temperature, and low density. In this case, the proton temperature within the ME is much higher than that preceding the shock, and thus also possesses a high plasma $\beta$ (not shown). Such features are indicative of a young ME at an early stage of its evolution, and a non-force-free magnetic field structure, which may have implications when fitting the magnetic field components with the typical force-free model solutions. \citet{lynch2022utility} previously explored the application of both force-free and non-force-free in situ flux rope models to flux ropes between 10 and 30 Rs, and found that both forms of model decently approximated the magnetic field structure. However, the usual caveats associated with the models still apply to flux ropes at such close distances to the Sun, perhaps even more so than when applied to observations at further heliocentric distances.

As discussed in \citet{romeo2023nearsun} and \citet{liu2024direct}, the ICME observations at PSP display more complex features. \citet{liu2024direct} describes how the magnetic field angle with respect to the radial direction ($\theta$) changes throughout the ME, likely indicating multiple structures within the ME. In addition, both \citet{romeo2023nearsun} and \cite{liu2024direct} identify the discontinuity in the radial magnetic field vector around 2022-09-06 17:30~UT to be a crossing of the HCS.

\begin{figure*}
\centering
{\includegraphics[width=\textwidth]{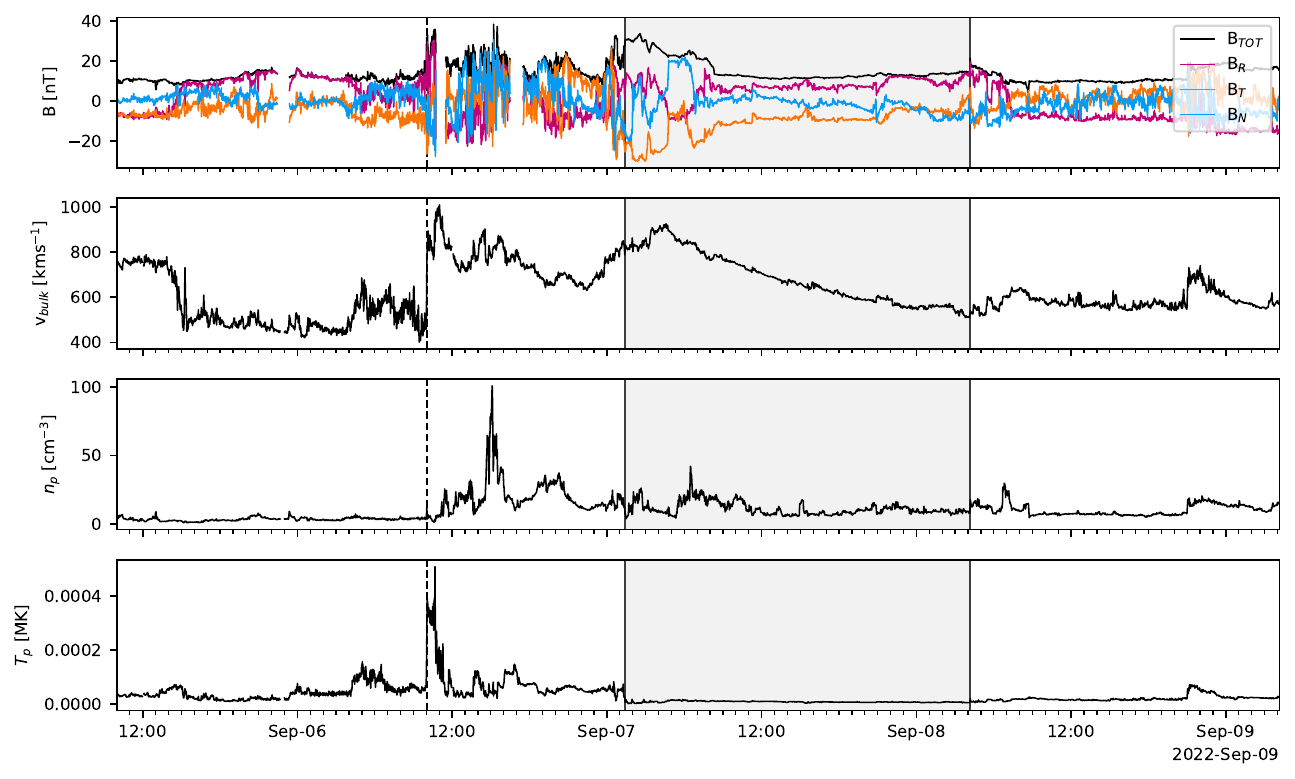}}
\caption{SolO in situ magnetic and plasma data, presented in the same format as Figure \ref{fig:psp_insitu}.} 
\label{fig:solo_insitu}
\end{figure*}

Figure \ref{fig:solo_insitu} presents the ICME magnetic field and plasma measurements observed by SolO, where we define the ICME start as the shock front identified at 2022-09-06 10:01~UT, coincident with a jump in magnetic field magnitude from $\sim$~12 to 32~nT and bulk proton velocity from 460 to 890\kms. The sheath region has a duration of approximately 15.4~hours, based on the enhanced proton temperature observations, and a largely varying magnetic field magnitude, possibly comprising many different structures. During the first 1.5~hours within the sheath, the proton temperature is very enhanced, around $\sim$4 MK. This structure is not coincident with a particularly large spike in proton density: this occurs later between approximately 2022-09-06 14:30 and 16:00~UT. Using the same technique as above, \citet{trotta2024properties} calculated a shock normal of $n = (0.6, -0.2, 0.7)$, thus $\theta = 51^\circ$ and the shock can be described as oblique, consistent with the result obtained at PSP.

At SolO, we define the start of the ME at 2022-09-07 01:25~UT, coincident with a large enhancement of the magnetic field strength, lower variation of the magnetic field components, and a drop in proton temperature. However, defining the ME boundary in this case is rather subjective, as it could also be defined as 2022-09-07 04:35~UT, coincident with the start of the declining bulk velocity profile also typically displayed by ICMEs.

From the magnetic field measurements, we are able to determine that the flux rope within the ME mostly rotates from North to South (from the start of the declining velocity profile), as shown by the positive to negative transition of the normal component (blue) after 2022-09-07 04:35~UT, whilst the transverse component (orange) remains pointing East (negative). We can therefore classify the ME flux rope as NES \citep{bothmer1998structure}: a right-handed flux rope, with a low inclination with respect to the ecliptic plane. The observations of the flux rope in situ are consistent with the analysis of the remote observations of its solar source, where \citet{long2023eruption} were able to deduce the right-handed rotation of the magnetic field from observations of the plasma flow along the filament channel.

Table \ref{tab:insitu_params} also lists the ICME parameters calculated within the ME at SolO. We calculated a mean magnetic field strength within the ME of 15.8~nT, with a maximum field strength of 33.8~nT. Assuming the ICME expands following $B \propto r^{-\alpha}$, we use the mean magnetic field strength with the ME to calculate the global expansion of the CME, resulting in a power of -1.74. \citet{salman2024survey} recently cataloged ICMEs observed by PSP prior to this event between 2018 October and 2022 August over a heliocentric distance range of 0.23 to 0.83~au, and found the relationship between mean magnetic field strength of the ME and heliocentric distance to have a power of  -1.21 $\pm$ 0.44. As discussed in \citet{salman2024survey}, the expansion rate is much lower than those found previously within 1~au e.g. -1.85 $\pm$ 0.07 for ICMEs observed by Helios 1 and 2 \citep{gulisano2010global}, -1.95 $\pm$ 0.19 for ICMEs observed by MESSENGER and ACE \citep{winslow2015interplanetary}, and -1.76 $\pm$ 0.04 for ICMEs observed by MESSENGER, Venus Express, STEREO A/B, and Wind \citep{good2019self}. The calculated power of -1.74 for the event in this study is in agreement with those studies, although outside of the range of \citet{salman2024survey}. 

Considering the proton velocity data, we are also able to compare the global expansion of the ICME to its local expansion at SolO. The dimensionless expansion parameter, $\zeta$, is a measure independent of radial distance, requiring only the cruise velocity, $V_c$ (i.e. speed at the mid-point of the flux rope), the rate of expansion, $\Delta V/\Delta t$, and heliocentric distance, $d$, of the spacecraft \citep[for more detail, see][]{demoulin2009causes, gulisano2010global}. \citet{gulisano2010global} found that $\zeta$ can be related to the global expansion of the ICME, where the magnetic field strength would be expected to decrease as $r^{-2\zeta}$ and radial size expected to increase as $r^\zeta$, therefore $\zeta$ = 0.5$\alpha$ as calculated above. For the velocity observations at SolO, if we take the start of the ME based on the start of the declining velocity profile i.e. at 2022-09-07 04:35~UT: $V_c$ = 640.6\kms, $\Delta V/\Delta t$ = 0.00396 km~s$^{-2}$, and $d$ = 0.682~au. Using these values, we find $\zeta$ = 0.925, and therefore expect the magnetic field strength to decrease with heliocentric distance as $r^{-1.85}$ and the size of the flux rope to scale as $r^{0.925}$. The resulting magnetic field relationship calculating using $\zeta$ is in fairly good agreement with that calculated using the observed mean values above (-1.74), which is often not the case as previous studies have found a weak correlation between measures of global and local expansion \citep{lugaz2020inconsistencies, davies2022multi}. To calculate the radial width of the ME at PSP and SolO, we simply multiply the duration by the cruise velocity, producing a value of 0.103~au at PSP and 0.375~au at SolO. Assuming a power law relationship as above, we find the size of the flux rope to scale as $r^{0.576}$. This power is much lower than the calculated $\zeta$, therefore, the dimensionless expansion parameter calculated using the velocity values at SolO, overestimates the rate of expansion of the ICME. This discrepancy may be due to a number of reasons: the different measurement tracks through the ICME taken by the spacecraft due to their longitudinal separation, the encounter with the flank rather than towards the nose of the ICME, or the different drag regimes due to very different heliocentric distances of the two spacecraft (discussed further in Section \ref{sec:ELEvoHI}). 

\section{Arrival time modeling: ELEvoHI} \label{sec:ELEvoHI}

\begin{figure*}
\centering
{\includegraphics[width=\textwidth]{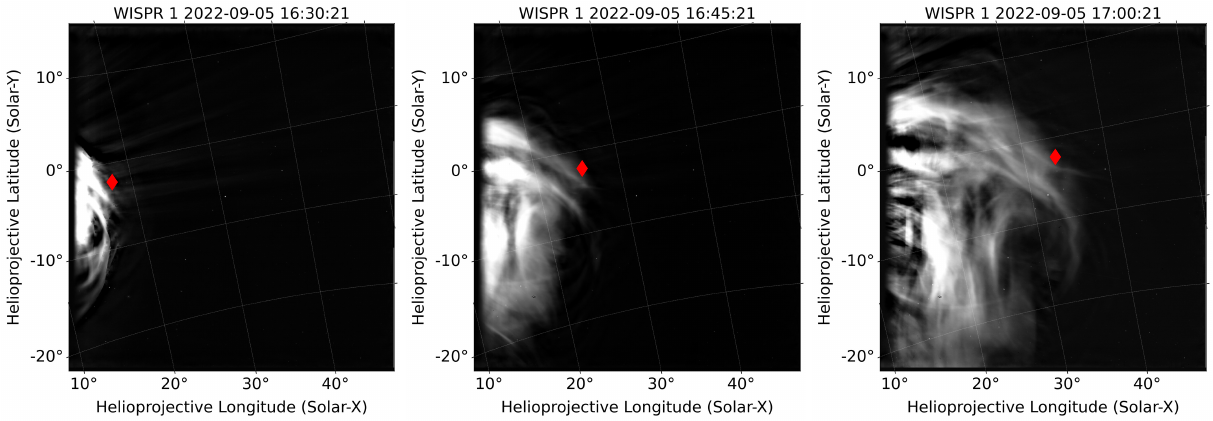}}
\caption{Consecutive images as PSP crossed the CME, observed by WISPR. The diamond-shaped marker indicates where the CME front was tracked along SolO's orbital plane.} 
\label{fig:wispr}
\end{figure*}

To demonstrate the connection between remote images and the observations in situ, we use the ELlipse Evolution model based on Heliospheric Imager data (ELEvoHI) introduced by \citet{rollett2016} to predict the arrival time of the CME at SolO and PSP. ELEvoHI combines the ELlipse Evolution model \citep[ELEvo;][]{moestl2015elevo} with drag-based model fitting \citep[DBM fitting;][]{vrsnak2013propagation} and can be run using an ensemble approach, as presented in \citet{amerstorfer2018ensemble} and \citet{amerstorfer2021evaluation}. It assumes a fixed propagation direction for an elliptical CME front whose evolution is governed by the solar wind drag. ELEvoHI requires the observed time-elongation profile of the event as input, which is obtained by tracking the leading edge of the CME in WISPR images, as shown in Figure \ref{fig:wispr}. We focus on the propagation along the plane in which SolO is located, which was quite close to that of PSP at the time of the event.

\begin{figure}
\centering
{\includegraphics[width=0.5\textwidth]{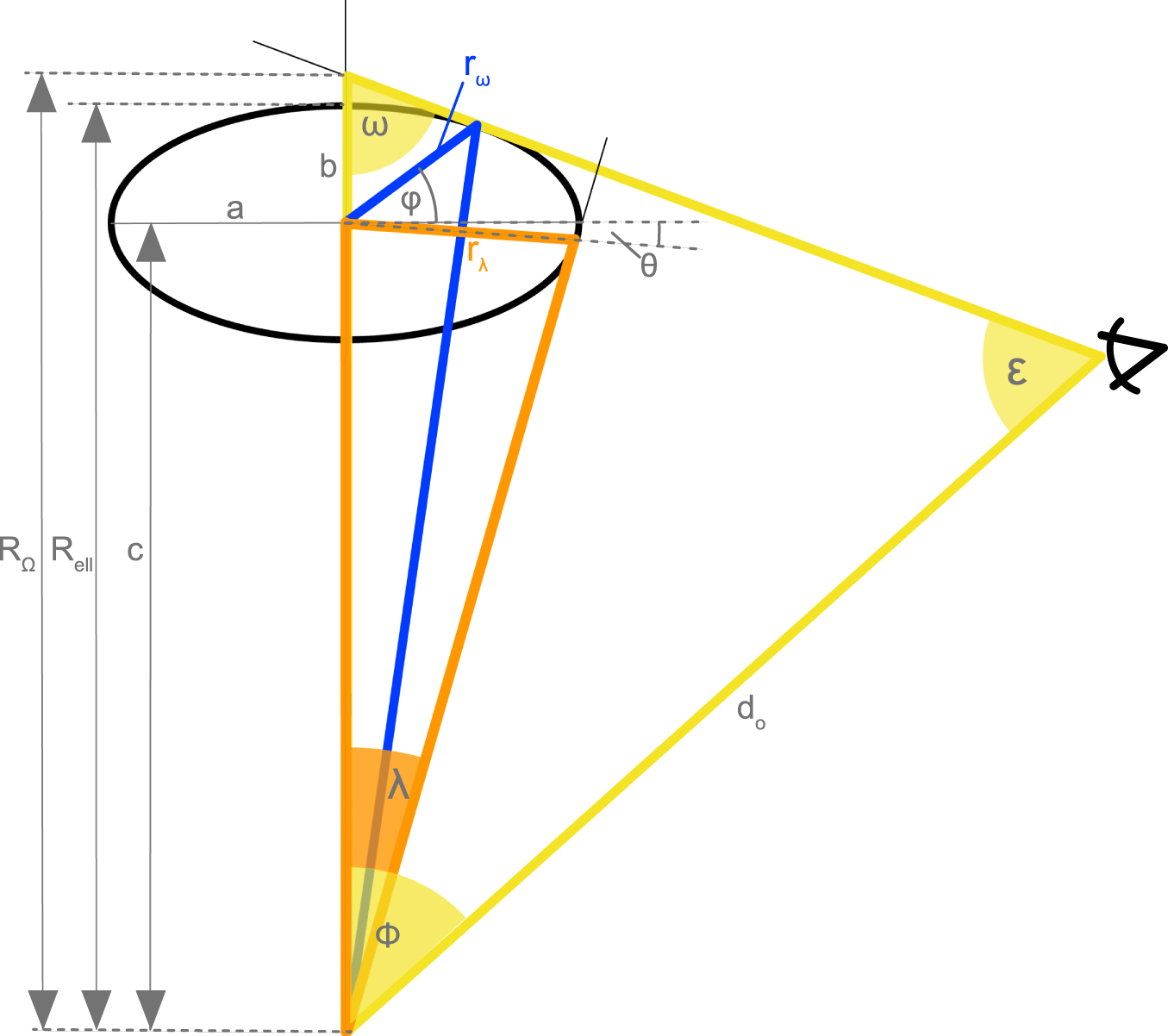}}
\caption{Overview of ELCon geometry, taken from \citet{rollett2016}.} 
\label{fig:elcon}
\end{figure}

Furthermore, ELEvoHI requires ranges of values for the inverse ellipse aspect ratio $f$, the direction of motion $\Phi$ of the CME apex relative to the observing spacecraft, the angular half-width $\lambda$, and the background solar wind speed $v_{sw}$. The parameters $f$, $\Phi$, and $\lambda$ are necessary inputs for the ELlipse Conversion (ELCon) method, which is used to transform the time-elongation into a time-distance profile, relying on the same assumptions about CME geometry that underpin ELEvoHI as a whole. The CME's elongation $\epsilon$ is the angle between the spacecraft's line-of-sight (LOS) and the Sun-spacecraft vector. ELCon treats the LOS as tangent to the CME's leading edge. A more detailed overview of the ELCon geometry can be found in Figure \ref{fig:elcon}, parameters from which will also be referenced in the following abbreviated derivation.

To derive the distance of the CME apex from the Sun, $R_{ell}$, it is necessary to calculate parameters $c$ and $b$:

\begin{equation}
R_{ell} = c + b
\label{eq:1}
\end{equation}

\begin{equation}
c=\frac{r_{\lambda} \sin(90+\theta - \lambda)}{\sin(\lambda)}
\label{eq:2}
\end{equation}

\begin{equation}
b=\frac{d_{\rm o} \sin(\epsilon) \sin(\lambda) \Omega_{\theta} \, \Omega_{\varphi}}{\sin(90+\theta-\lambda) \sin(\omega) \Omega_{\varphi}+\sin(90+\varphi-\omega) \sin(\lambda)\Omega_{\theta}}
\label{eq:3}
\end{equation}

with

\begin{equation}
\Omega_x =\sqrt{f^2 \cos^2 x + \sin^2 x},  \quad x \in \{\theta, \varphi \}.
\label{eq:4}
\end{equation}

To obtain the distance from the Sun at any point along the CME front, it is necessary to calculate the off-axis correction, which considers the angular offset $\Delta$ between the direction of the CME apex and the location of interest:

\begin{equation}
 R_{is}=\frac{c \cos\Delta + \sqrt{(b^2-c^2) f^2 \sin^2\Delta+b^2 \cos^2\Delta}}{f^2 \sin^2\Delta+\cos^2\Delta}.
 \label{eq:5}
 \end{equation}

ELCon has so far been exclusively applied to analyze events observed by the HI instruments onboard the STEREO twin-spacecraft, which orbit the Sun at a distance of around 1~au. As demonstrated by \citet{patel2023wispr}, the assumptions used to convert elongations into distances are non-trivial when working with WISPR data and can significantly influence the reconstructed CME kinematics. \citet{nindos2021wispr} proposed a new approach, so-called R-Maps, based on the assumption that the signal of each pixel originates at the intersection between the LOS and the Thomson surface (\citealt{vourlidashoward2006thomson}).

While the basic geometric principles governing ELCon remain valid independent of the observer's distance to the Sun, complications may arise due to the fast motion of the spacecraft and the observer-directed nature of the event. Rapid changes in the observer's location could result in different parts of the CME being tracked, distorting results. If the observer enters the CME, inferring kinematics using ELCon is no longer possible due to the initial assumption of the LOS being tangent to the front becoming unreasonable. The track used for our prediction covers a time span of around one hour, from 16:30~UT to 17:35~UT. The in-situ arrival time of the CME at PSP was determined to be 17:27~UT, meaning our assumptions are valid for most of the time-elongation profile. The limited duration of the track also minimizes the impact of spacecraft motion on CME reconstruction. Comparison between distances obtained using ELCon and R-Maps seems to indicate that ELCon yields plausible results for this time period, but further study is needed to fully gauge to what extent ELCon is applicable to different observing geometries.

After using ELCon, we fit a drag-based equation of motion \citep[][]{vrsnak2013propagation} to the derived CME kinematics. In that way we gain the drag-parameter, $\gamma$, and the ambient solar wind speed, $v_{sw}$, which are further used for modeling the CME evolution using the last part of the ELEvoHI model suite, ELEvo \citep[][]{moestl2015elevo}.

Running ELEvoHI in ensemble mode allows us to explore different combinations of input parameters, giving us an estimate of the modeling uncertainty. For this study, the values for $\Phi$ ranged from 0 to 60$^{\circ}$ and were chosen to cover the range of possible values modeled by 3DCORE (see Section~\ref{sec:3DCORE}), as well as the location of the source region. $\lambda$ was specified based on estimates from the DONKI catalog, resulting in a variation between 40 and 80$^{\circ}$. We varied the solar wind speed between 250 and 700~km~s$^{-1}$ and the inverse ellipse aspect ratio from 0.7 to 1.0.

\begin{figure}
\centering
{\includegraphics[width=0.7\textwidth]{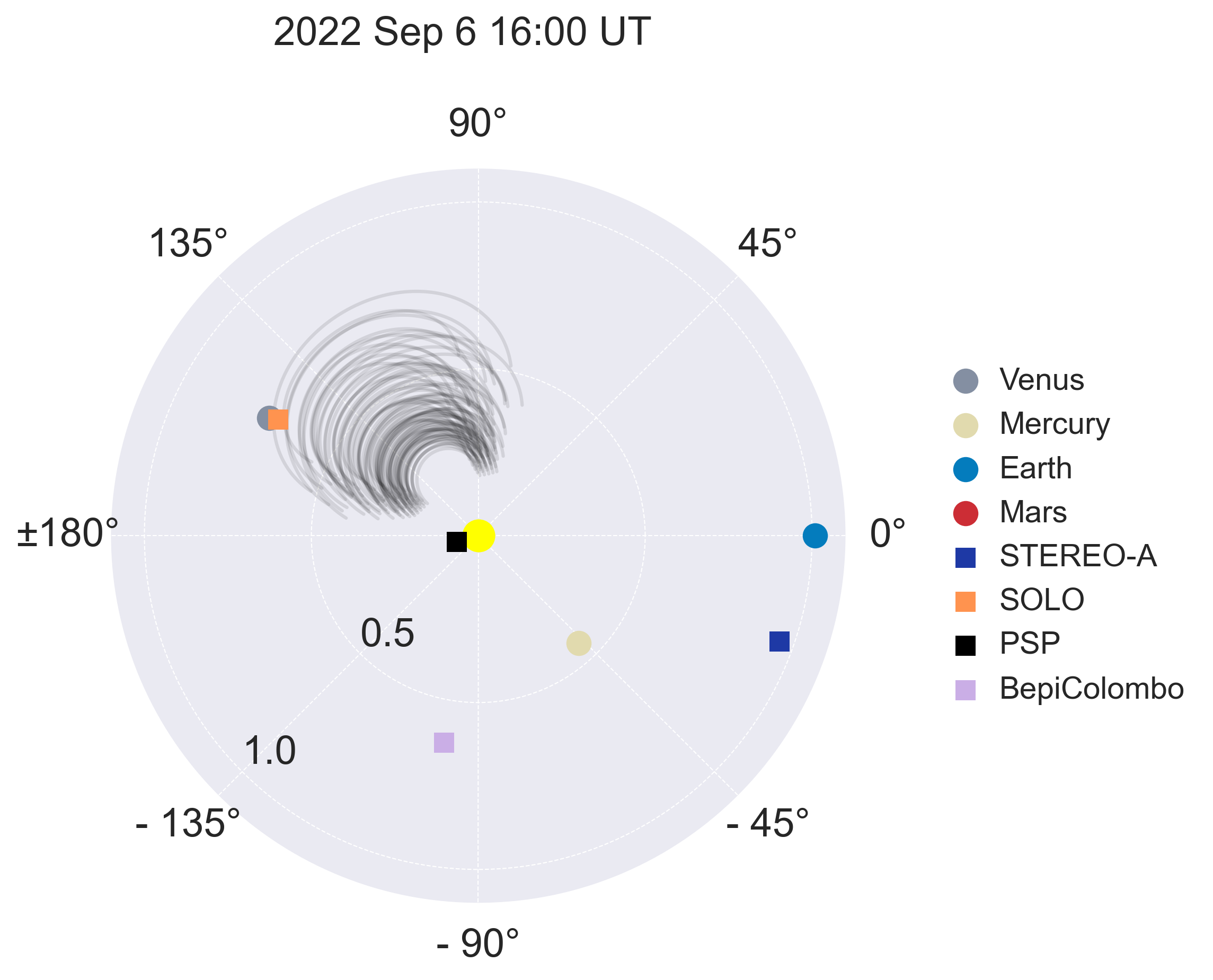}}
\caption{Results of the ELEvoHI model ensemble runs based on WISPR data for the time 2022 Sep 6 16:00~UT in HEEQ coordinates, 6 hours after the observed shock arrival at SolO. Different CME fronts represent the different ensemble members produced by varying the input parameters. SolO is represented by the orange marker, and PSP by the black marker.} 
\label{fig:elevohi}
\end{figure}

Figure \ref{fig:elevohi} shows the propagation of the CME as modeled by ELEvoHI for all ensemble members at a particular timestamp (2022 Sep 6 16:00~UT). The 468 ensemble runs yield a mean average arrival error of 0.3~hours for PSP and 18.4~hours for SolO. This discrepancy between the modeled arrival times suggests that the underlying model assumptions may not be suitable for making predictions based on time-elongation tracks obtained from close proximity to the Sun. Usually, DBM fitting is done at distances greater than 20~$R_{\odot}$ from the Sun, as it has been shown that the drag force becomes dominant while the influence of the Lorentz force becomes increasingly negligible for most cases at this distance \citep{vrsnak2010drag,vrsnak2013propagation}. However, in our study, the CME is tracked much closer to the Sun, resulting in the time-distance profile only extending to 14~$R_{\odot}$. The drag regime at larger distances is markedly different from that closer to the Sun, where additional factors can significantly alter the expected trajectory and arrival time of the CME. Furthermore, the solar wind speed derived via fitting is assumed to stay constant during CME propagation, which is not usually the case. For most ensemble members an ambient solar wind speed of 700 km~s$^{-1}$ was found based on the CME kinematics in the early propagation phase, where the propagation is not governed by the drag-force alone, as mentioned above. Additionally, it is possible that effects related to observing geometry introduce distortions in the reconstruction of CME kinematics using ELCon \citep[e.g.][]{hinterreiter2021elevohi, hinterreiter2021belevohi}.

A second CME was observed by SOHO LASCO/C2 on 2022 Sep 5 18:39~UT. Remnants of the preceding CME largely obscured it, resulting in only faint visibility in LASCO/C2 and STEREO-A/COR2 images. Estimates using coronagraph images by NASA's Moon to Mars Space Weather Analysis Office placed the CME at a longitude of -144$^{\circ}$, making it a far-sided event. Simulations carried out using the WSA-ENLIL+Cone model (\citealt{argepizzo2000improvement}, \citealt{odstrcil2003enlil}, \citealt{arge2004stream}, \citealt{odstrcil2004numerical}) showed that it is unlikely to have impacted SolO or PSP (see \url{https://kauai.ccmc.gsfc.nasa.gov/DONKI/view/CMEAnalysis/21563/1}), leaving the first CME as the one most likely to have been observed in situ at both spacecraft.

However, it cannot be ruled out that the second event interacted with the preceding CME, impacting its propagation. Considering this, as well as the variability in solar wind conditions at the time the events took place, it is likely that the initial CME experienced some degree of deformation, further complicating the modeling when using ELEvoHI's assumption of a rigid elliptical front.

\section{Flux rope modeling: 3DCORE} \label{sec:3DCORE}

In order to reconstruct the parameters of the flux rope of the ICME observed by PSP and SolO, we use the semi-empirical 3D Coronal Rope Ejection (3DCORE) model, which was introduced by \cite{moestl20183dcore}, and further developed by \cite{weiss2021triple} and \cite{weiss2021analysis}. The version used in this paper is thoroughly described in \citet{ruedisser2024understanding} (ApJ, in revision).

3DCORE assumes a Gold-Hoyle-like magnetic field embedded within a toroidal flux rope with an elliptical cross-section that expands self-similarly while the both ends of the flux rope remain attached to the Sun. The propagation of the 3DCORE flux rope outwards through the heliosphere is considered by using a drag-based model \citep{vrsnak2013propagation}. 3DCORE can be used to infer flux rope parameters by fitting in situ magnetic field observations, using an Approximate Bayesian Computation - Sequential Monte Carlo algorithm \cite[first introduced for 3DCORE by][]{weiss2021analysis}. This algorithm utilizes the model's advantage of being able to produce mega ensembles on the order of $10^6$ runs per second, to search the input parameter space for possible fits to the in situ data. By accordingly restricting the input parameter range, priors from either auxiliary measurements or physical considerations can be incorporated. 

During the fitting process, the model is evaluated at a given set of fitting points and compared to the measured in situ data. Model runs that produce a non-zero magnetic field at the boundaries of the event are rejected, as well as runs that do not produce signatures at the fitting points. This leads to a restriction of the input parameters to combinations that generate a ``hit'' at the observing spacecraft. Each iteration of the fitting process then serves to further confine the input parameters to minimize the root mean square error (RMSE) between model values at the fitting points and the corresponding in situ measurements. This procedure finally yields an ensemble of accepted combinations of input parameters.

3DCORE is able to simultaneously fit multiple magnetic field measurements from different spacecraft \citep{weiss2021triple}, however, this only works well when the spacecraft are relatively close together. In the case of larger separations, the general assumptions of 3DCORE, such as a constant background solar wind speed, cannot be modeled through a single value that complies with both measurements. Due to the large radial separation of PSP and SolO for the event under study ($\sim$~0.625~au), we apply 3DCORE to both observations separately.

3DCORE also has the ability to take into account the motion of the spacecraft, which is especially useful considering the faster movement of PSP in comparison to other spacecraft at further heliocentric distances. The spacecraft motion is considered by fitting the magnetic field observations in the HEEQ coordinate frame. We use the same version 3DCORE to fit both the PSP and SolO magnetic field observations, although we expect the spacecraft motion of SolO to have little effect on the fitting results in comparison to those at PSP.

\begin{deluxetable}{ccccccccc}
\tabletypesize{\scriptsize}
\tablecaption{Parameter ranges and the corresponding results of the 3DCORE fitting for SolO and PSP. \label{tab:combined}}
\tablehead{
\colhead{} & \colhead{} & \multicolumn{3}{c}{SolO} & \colhead{} & \multicolumn{3}{c}{PSP}\\
\cline{3-5}
\cline{7-9}
\colhead{Parameters} & \colhead{Units} & \colhead{Minimum} & \colhead{Maximum} & \colhead{Solution} & \colhead{} & \colhead{Minimum} & \colhead{Maximum} & \colhead{Solution} \\
}
\startdata
longitude & deg & $100.$ & $200.$ & $128.5 \pm 5.95$ &  & $110.$ & $170.$ & $135.11 \pm 11.69 $ \\
latitude & deg & $-40.$ & $40.$ & $-5.86 \pm 7.01$ & & $-40.$ & $40.$ & $5.85 \pm 10.11 $ \\
inclination & deg & $-45.$ & $45.$ & $5.61 \pm 6.16$ & & $0.$ & $60.$ & $20.99 \pm 12.39 $ \\
d$_{1\mathrm{AU}}$ & au & $0.2$ & $0.35$ & $0.34 \pm 0.01$ & & $0.25$ & $0.45$ & $0.42 \pm 0.02 $ \\
$\delta$ &  & $1.$ & $4.$ & $2.78 \pm 0.77$ & & $1.$ & $5.$ & $3.43 \pm 1.04 $ \\
T$_{\mathrm{fac}}$ & & $0.$ & $200.$ & $9.7 \pm 2.87$ & & $0.$ & $200.$ & $17.68 \pm 10.32 $ \\
$v_{0}$ &  km~s$^{-1}$ & $500.$ & $1900.$ & $1619.28 \pm 173.09$ & & $200.$ & $1500.$ & $318.79 \pm 87.93 $ \\
B$_{1\mathrm{AU}}$ & nT & $5.$ & $20.$ & $10.64 \pm 1.23$ & & $5.$ & $20.$ & $12.6 \pm 4.06 $ \\
$\Gamma$ &  & $0.2$ & $3.$ & $0.39 \pm 0.12$ & & $0.2$ & $3.$ & $1.69 \pm 0.76 $ \\
$v_{\mathrm{sw}}$ &  km~s$^{-1}$ & $100.$ & $900.$ & $440.32 \pm 70.86$ & & $100.$ & $700.$ & $267.79 \pm 119.57 $ \\
$n_{a}$ &  & $1.$ & $1.3$ & $1.17 \pm 0.08$ & & $1.$ & $1.3$ &  $1.04 \pm 0.03 $ \\
$n_{b}$ & & $1.$ & $2.$ & $1.76 \pm 0.18$ & & $1.2$ & $2.$ & $1.47 \pm 0.15 $\\
\enddata
\tablecomments{The launch radius $r_{0}$ is fixed to be $10.7$~$R_{\odot}$. The propagation longitude and latitude of the flux rope are given in HEEQ coordinates, and the inclination (orientation of the axis of the flux rope), width of cross-section at 1~au (d$_{1\mathrm{AU}}$), aspect ratio of cross-section at 1~au ($\delta$), a twist factor (T$_{\mathrm{fac}}$) which is related to the number of twists over the whole torus, launch speed ($v_0$), launch radius ($r_0$), magnetic field strength at 1~au (B$_{1\mathrm{AU}}$), drag parameter ($\Gamma$), global solar wind speed ($v_{\mathrm{sw}}$), expansion rate ($n_a$) of the cross-section diameter, and the rate at which the magnetic field decreases ($n_b$) are listed.}
\end{deluxetable}

Table \ref{tab:combined} presents the parameter ranges chosen for both the SolO and PSP fitting. 3DCORE takes the following input parameters: propagation direction (longitude and latitude) in HEEQ coordinates, orientation of the axis of the flux rope (inclination), width of cross-section at 1~au (d$_{1\mathrm{AU}}$), aspect ratio of cross-section at 1~au ($\delta$), a twist factor (T$_{\mathrm{fac}}$) which is related to the number of twists over the whole torus, launch speed ($v_0$), launch radius ($r_0$), magnetic field strength at 1~au (B$_{1\mathrm{AU}}$), drag parameter ($\Gamma$), global solar wind speed ($v_{\mathrm{sw}}$), expansion rate ($n_a$) of the cross-section diameter, and the rate at which the magnetic field decreases ($n_b$). The launch time was chosen as 2022 Sep 5 16:00~UT at a launch radius $r_{0}$ of $10.7$~$R_{\odot}$, to coincide with WISPR observations. Choosing the input parameter ranges, we attempt to include physical knowledge gained from additional sources and physical considerations, such as flux rope type (low inclination and positive twist factor) or spacecraft position restricting the propagation direction (longitude and latitude). This approach ensures timely convergence of the model, while still keeping the remaining model parameters rather free for the algorithm to optimize.

The results obtained for the SolO fit were then used to restrict the input parameter ranges for the PSP fit. Additionally, allowing higher values for d$_{1\mathrm{AU}}$ and $\delta$, as well as an unusually low $v_0$ due to the extremely close distance of PSP to the Sun, increased the possibility of generating hits in the first place. This approach was chosen as 3DCORE appeared to have difficulties converging to a minimum. Due to PSP being very close to the Sun as well as its rapidly changing position, a highly varying range of input parameters can be used to generate more or less equally ``good'' signatures. This behavior was somewhat expected, as even characteristics such as flux rope type were difficult to establish visually. While these circumstances hinder the inference of flux rope parameters through solely utilizing the PSP measurements, we can still test which of the assumptions from SolO hold for the PSP fit. 

\begin{figure*}
\centering
{\includegraphics[width=\textwidth]{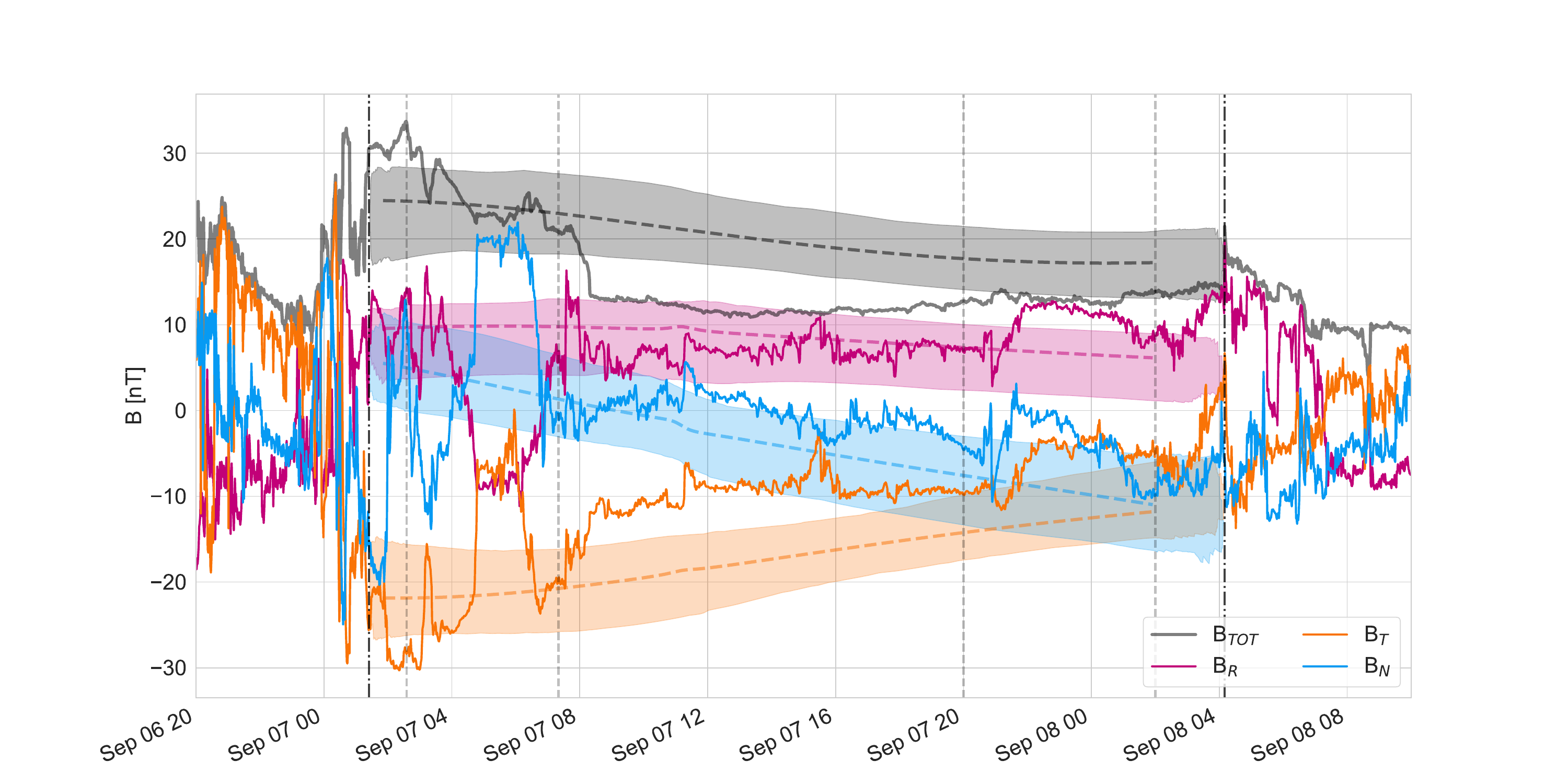}}
\caption{%
SolO magnetic field measurements (solid lines) together with the 3DCORE flux rope ensemble. The shaded area corresponds to the 2$\sigma$ spread of the ensemble. The dashed colored lines show one specific flux rope from the ensemble. The vertical black dashed lines denote the beginning and the end of the reconstruction interval used by 3DCORE. The vertical grey dashed lines denote the fitting points used for comparing the model output to the in situ data. The magnetic field measurements are in local spacecraft RTN coordinates.
}%
\label{fig:3dcore_solo}
\end{figure*}

Figure \ref{fig:3dcore_solo} presents the reconstructed 3DCORE flux rope (dashed lines) together with the in situ magnetic field observations of SolO (solid lines), where the shaded area corresponds to the 2$\sigma$ spread of the ensemble. The vertical grey dashed lines throughout the defined ME indicate the fitting points used for comparing the model output to the in situ data. As discussed in Section \ref{subsec:insitu_obs}, the observations display a magnetic field rotation that indicates a low-inclination right-handed NES flux rope type. As shown in Table \ref{tab:combined}, we restrict the inclination between -45 and 45\dg to reflect a low inclination flux rope, and restrict the twist factor to positive values to align with the in situ observations. For the case of the SolO observations, 3DCORE was able to reconstruct the flux rope without too many adjustments of parameter fitting limits and in a short computational time.

\begin{figure*}
\centering
{\includegraphics[width=\textwidth]{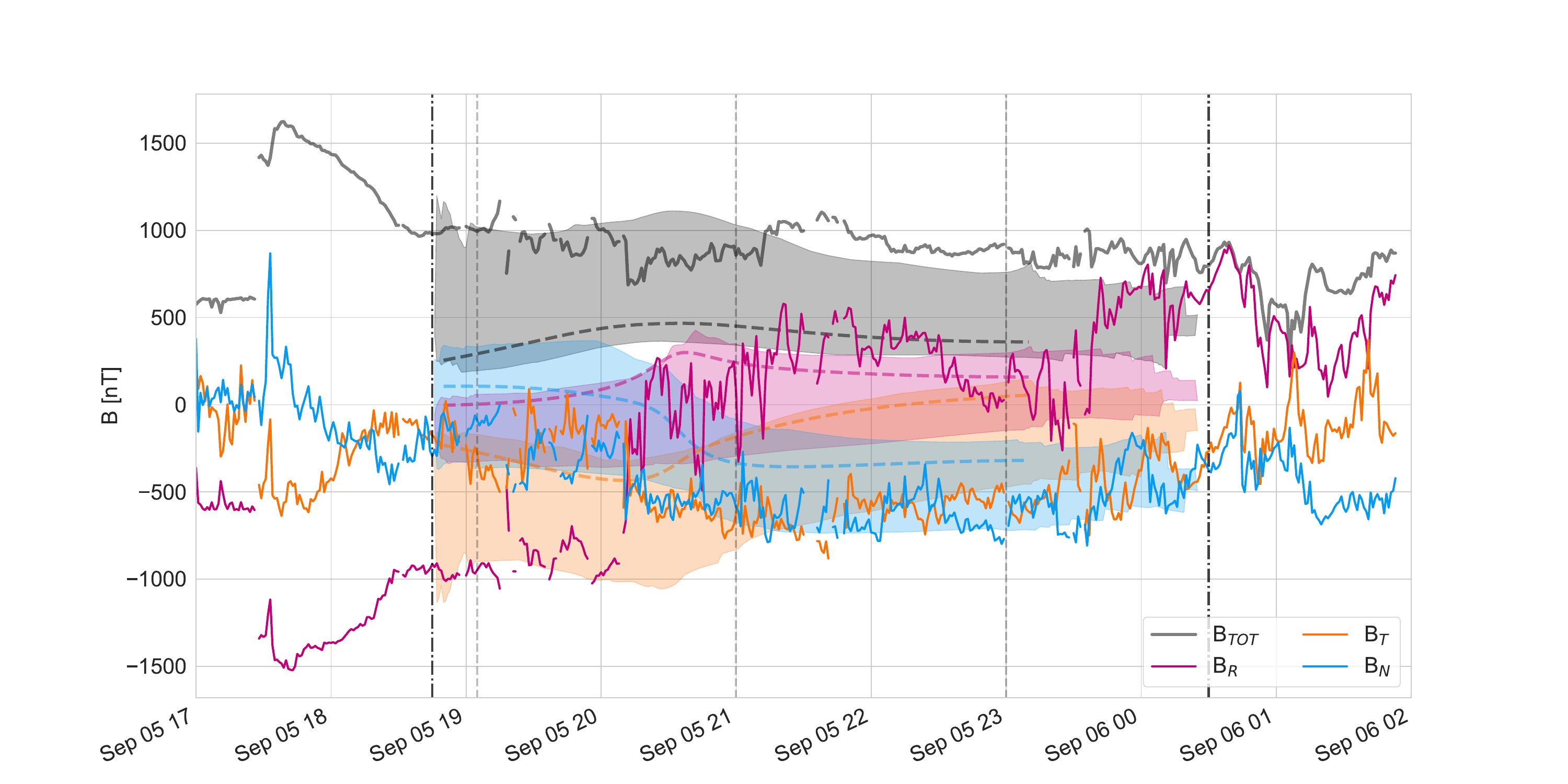}}
\caption{
PSP magnetic field measurements (solid lines) together with the 3DCORE flux rope ensemble, presented in the same format as Figure \ref{fig:3dcore_solo}.
} 
\label{fig:3dcore_psp}
\end{figure*}

Similarly, Figure~\ref{fig:3dcore_psp} shows the PSP in situ magnetic field measurements with the flux rope ensemble generated by 3DCORE. Despite the flux rope type being difficult to establish visually from the in situ observations, 3DCORE fits the PSP data relatively well when restricted to positive twist values, consistent with the SolO in situ observations of a right-handed flux rope type, as well as with the inferred magnetic field rotation due to the plasma flows observed along the solar source of the flux rope \citep{long2023eruption}. The fitting captures the overall trend of the in situ observations quite well, however, we see some substantial deviation especially in the B$_R$ component, which exhibits strong variations hinting at additional processes affecting the flux rope, as discussed in \citet{romeo2023nearsun} and \citet{liu2024direct}.

Similar 3DCORE fitting was performed using the PSP in situ observations in \citet{long2023eruption}. We note that in \citet{long2023eruption}, the model assumed a stationary spacecraft when fitting the data. Although the assumption of a stationary spacecraft holds at larger distances from the Sun, in this study, we improve on the fitting to capture the fast movement of PSP relative to the motion of the flux rope over the spacecraft.

Table~\ref{tab:combined} also presents the flux rope parameters inferred by 3DCORE for the observations at PSP and SolO. Comparing the values in Table \ref{tab:combined}, we can see that even some of the less restricted parameters are similar for both reconstructions, for example, the propagation longitude and the scaled mean magnetic field strength at 1~au, B$_{1\mathrm{AU}}$. The modeled launch speed of 1619.28 $\pm$ 173.09~km~s$^{-1}$ for SolO is comparable with the mean shock speed of 1520~km~s$^{-1}$ calculated in \citet{trotta2024properties}. The non-physically low launch speed of 318.79 $\pm$ 87.93~km~s$^{-1}$ for the PSP reconstruction hints towards the models inability to accurately reproduce the CME evolution close to the Sun.

Larger differences are seen in the propagation latitude, the inclination, $\delta$, T$_\mathrm{fac}$, $\Gamma$, and $v_{\mathrm{sw}}$. The propagation latitude of -5.86 $\pm$ 7.01\dg produced at SolO, indicates a southwards propagation (albeit, very slightly) with respect to the Earth-equatorial plane, in agreement with the southwards propagation observed in the coronagraphs and indicated by the location of the solar source, and the GCS fitting results of R.~Patel \citep{patel2023wispr} mentioned in Section \ref{subsec:solar_source}. The PSP value of 5.85 $\pm$ 10.11\dg suggests a propagation direction slightly above the solar equatorial plane, however, the two results lie within the calculated 2$\sigma$ spread of the ensemble members. These larger uncertainties are likely produced due to the way in which PSP crosses the flank of the ICME. 

The inclination value produced by 3DCORE at SolO is 5.61 $\pm$ 6.16\dgg, in agreement with the low inclination flux rope type determined from observations. At PSP, 3DCORE produces a value of 20.99 $\pm$ 12.39\dgg, a more intermediate flux rope inclination than that of SolO. As the inclination value can be used to set the flux rope type for a 3DCORE model \citep{ruedisser2024understanding}, the high uncertainty regarding the inclination at PSP reflects the ambiguity in visually determining the flux rope type from the PSP in situ observations. The higher inclination at PSP is also not surprising given the estimated inclination of 58\dg from the GCS fitting results using the STEREO-A/COR-2 images. The declining inclination between coronagraph observations, PSP, and SolO, is in agreement with previous suggestions that ICME magnetic flux ropes may align with the HCS as they propagate \citep{yurchyshyn2008relationship, isavnin2014three}, and multi-spacecraft studies which found small decreases in flux rope axis inclination as ICMEs propagated to larger heliocentric distances \citep{good2019self, davies2022multi}. 

The magnetic field expansion rates, $n_b$, produced by 3DCORE are -1.47 $\pm$ 0.15 at PSP and -1.76 $\pm$ 0.18 at SolO. These values are directly comparable to the magnetic expansion rate of -1.74 calculated using the in situ observations in Section \ref{subsec:insitu_obs}. The magnetic field expansion rate for SolO is in very good agreement with that calculated using the in situ magnetic field observations at both spacecraft. The rate at PSP is much lower, however, it is within the uncertainty range of -1.21 $\pm$ 0.44, calculated using events prior to this event observed by PSP in \citet{salman2024survey}.

Similarly to the magnetic field expansion rate, we are also able to compare the expansion rate of the cross-section diameter produced by 3DCORE to previous calculations based on the in situ observations in Section \ref{subsec:insitu_obs}. There, we found the size of the flux rope expected to scale as $r^{0.576}$. In comparison, the expansion rates produced by the 3DCORE fitting are 1.04 $\pm$ 0.03  and 1.17 $\pm$ 0.08 at PSP and SolO, respectively: much greater than that observed. Interestingly, we find that the expansion rate value at PSP is actually more similar to the value of the dimensionless expansion parameter, $\zeta$ = 0.925, calculated in Section \ref{subsec:insitu_obs} using the plasma velocities observed at SolO.

At both SolO and PSP, the estimated cross-section at 1~au produced by the model, d$_{1\mathrm{AU}}$, is toward the larger side of ICMEs at 1~au \citep[typically around 0.2~au;][]{lepping1990magnetic} and should be taken with caution. The scaling law used in 3DCORE is also most probably not applicable for every ICME throughout the whole Sun to spacecraft propagation. As discussed in Section \ref{sec:ELEvoHI}, the drag regime at larger distances is very different from that closer to the Sun, and we expect this to affect the expansion rate; at PSP, it is clear that the expansion is overestimated by 3DCORE.

\begin{figure*}
\centering{\includegraphics[width=.95\textwidth]{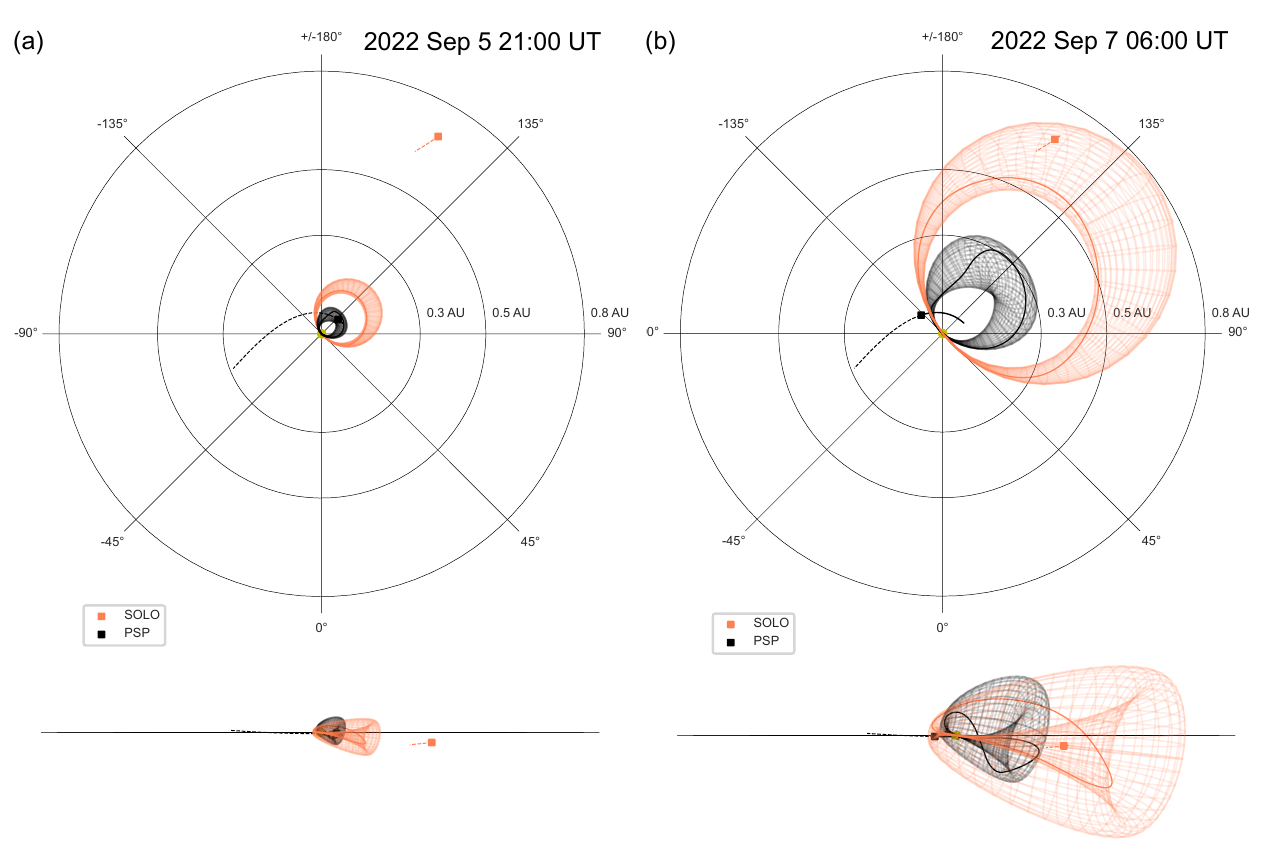}}
\caption{Representative flux ropes reconstructed by 3DCORE from the PSP (black) and SolO (orange) observations, (a) at a time of observation of the magnetic ejecta by PSP (2022 Sep 5 21:00~UT) and (b) at a time of observation the magnetic ejecta at SolO (2022 Sep 7 06:00~UT). The solid lines inside of the flux rope grids depict a magnetic field line and its twist. Top: top view down from solar equatorial north, bottom: side view along the solar equatorial plane. The black solid and dashed line shows the trajectory of PSP before and after the observation time; the orange dashed line is the trajectory of SolO after the observation.}
\label{fig:3d_reconstruction}
\end{figure*}

3DCORE provides only one global background solar wind speed, and due to the large radial separation of the two spacecraft, the background solar wind speed, $v_{\mathrm{sw}}$, is very different for the two fits: 267.79 $\pm$ 119.57 km~s$^{-1}$ at PSP, and 440.32 $\pm$ 70.86 km~s$^{-1}$ at SolO. Comparing these to the mean measured in situ background solar wind speeds at the two spacecraft (calculated over the preceding 12 hours before the ICME arrival time), we find a mean background solar wind speed of 429.61 km~s$^{-1}$ at PSP and 504.93 km~s$^{-1}$ at SolO. The modeled $v_{\mathrm{sw}}$ is much more similar to the observed mean value at SolO, whereas at PSP, the modelled $v_{\mathrm{sw}}$ is 60\% of the observed mean value. Such a low modeled $v_{\mathrm{sw}}$ is a product of the 3DCORE fitting procedure at PSP, where the model was more likely to find hits at lower speeds so close to the Sun, as is also seen in the very low launch velocity obtained by 3DCORE. 3DCORE was mainly developed to reconstruct flux rope signatures further out from the Sun, thus it has more difficulty when fitting the in situ signature at PSP so close to the Sun. The small modeled background solar wind speed for the PSP flux rope therefore results in a stronger deceleration compared to the SolO flux rope. Together with the already low launch speed, this leads to a very slow propagation speed for the PSP flux rope. This is clearly demonstrated in Figure \ref{fig:3d_reconstruction} which presents the 3D reconstructions of the global structure using the 3DCORE parameters at SolO (represented by the orange grid) and PSP (black grid), where views from top down and along the solar equatorial plane are shown. Figure \ref{fig:3d_reconstruction}a presents the reconstructed flux ropes around the time of observation at PSP (2022 Sep 5 21:00~UT) and Figure \ref{fig:3d_reconstruction}b presents the reconstructed flux ropes around the time of observation at SolO (2022 Sep 7 06:00~UT), one day and 9 hours later, giving a sense of the time evolution of the flux rope as it propagated. The differences between the arrival times are not as distinct at the earlier PSP observation time, however, the difference in timing is very clear at the later SolO observation time, highlighting the large difference in deceleration produced due to the background solar wind speeds. 

\section{Summary and Conclusions} \label{sec:conclusion}

The CME launched on 2022 September 5 is the closest to the Sun detected in situ to date, observed by PSP at a distance of only 0.07~au. Due to a longitudinal separation within 30\dgg, the same CME was also observed in situ by SolO at a heliocentric distance of 0.69~au. Such a spacecraft configuration has provided an excellent opportunity to investigate how the flux rope of the CME evolved between a very early stage of its evolution, out to heliocentric distances much further away.

In this study, we used both in situ observations at PSP and SolO to determine its radial evolution, as well as two models: ELEvoHI, which provided a direct link between remote images of the CME at PSP to in situ arrival times of the ICME, and 3DCORE, a semi-empirical 3D flux rope model, capable of inferring flux rope parameters by fitting in situ magnetic field observations and reconstructing the global 3D structure of the CME.

The in situ observations at PSP appear very different to those at SolO. The strong rotation in the radial component of the magnetic field observed by PSP, implies that the spacecraft crossed through the flux rope at an angle to the solar equatorial plane \citep[e.g. see Figure 2 of][for examples of different trajectories through the flux rope cross section]{dibraccio2015messenger}. This situation is reflected in the changing latitude of PSP as the ICME passed over the spacecraft (see Table \ref{tab:insitu_params}), and the southwards propagation of the CME itself. Such observations make determining the flux rope type, and thus handedness and inclination, much trickier. At SolO, the magnetic field components are observed to rotate from North-East-South, and thus we are able to determine that the flux rope was of low-inclination with a right-handed rotation. This right-handed rotation is in agreement with the right-handed motion of plasma flows along its solar source filament channel \citep{long2023eruption} and the right-handed nature of the active region. 

Observations of the solar source of the CME and results from GCS fitting show that the CME propagated in a southwards direction with respect to the solar equatorial plane, and thus PSP and SolO likely observe the flank of the CME. This interpretation is consistent with the in situ observations, where \citet{trotta2024properties} determined the shock front to be oblique. Despite the flank encounter, the global expansion relationship calculated using observations of the mean magnetic field within the ME, $B \propto r^{-1.74}$, is surprisingly consistent with that determined by calculating the dimensionless expansion parameter (a measure of local expansion) at SolO, $B \propto r^{-1.85}$. However, similar scaling relationships between the size of the magnetic flux rope determined from in situ observations, $D \propto r^{0.576}$, and that of the dimensionless expansion parameter, $D \propto r^{0.925}$, are not consistent, where the dimensionless expansion parameter significantly overestimates the expansion rate.

To demonstrate the connection between remote images and the observations in situ, we use the ELEvoHI model to predict the arrival time of the CME at SolO and PSP. In this case, we use a time-elongation profile obtained by tracking the leading edge of the CME in WISPR images as an input. We find a large discrepancy between modeled arrival times at PSP and SolO: a mean arrival error of 0.3~hours and 18.4~hours, respectively. These results suggest that the underlying model assumptions may not be suitable for making predictions based on time-elongation tracks obtained from close proximity to the Sun (up to only 14~$R_{\odot}$), where the drag regime is markedly different in comparison to larger heliocentric distances.

The 3DCORE model is capable of fitting the magnetic field observations at each spacecraft to estimate parameters such as the propagation direction, inclination, and expansion rate of the CME flux rope. Due to the large radial separation of PSP and SolO, we applied 3DCORE to both observations separately, including physical knowledge gained from additional sources and physical considerations, and taking the results of the SolO fit to restrict the input parameters for the PSP fit. We find that although many parameter results are similar between the two fits, it is difficult to reconstruct a completely consistent global 3D structure between the in situ observations at PSP and SolO. This is likely due to the single spacecraft trajectories through the CME, with large radial and longitudinal separations, as features often differ throughout the flux rope and are perhaps more localized to the in situ measurement site. When reconstructing and propagating the ICME, the difference between the background solar wind speeds produced by 3DCORE for PSP and SolO were strongly apparent, with the low speed at PSP decelerating the CME significantly. From our model fitting results, it is clear the solar wind background and drag regime strongly affects the modeled expansion and propagation of CMEs and needs to be considered. 

These results underscore the importance of refining our models and questioning the validity of our assumptions for CME propagation as we seek to improve the accuracy of space weather forecasting tools. As PSP and SolO continue their missions, validation of such models with in situ multi-spacecraft observations is crucial towards creating a more realistic picture of CME evolution across large heliocentric distance ranges and differing solar wind regimes, and incorporated into models continuing to improve our understanding of CME propagation throughout the heliosphere. 


\begin{acknowledgments}

E.~D., U.V.~A., H.T.~R., C.~M., and E.~W. are funded by the European Union (ERC, HELIO4CAST, 101042188). Views and opinions expressed are however those of the author(s) only and do not necessarily reflect those of the European Union or the European Research Council Executive Agency. Neither the European Union nor the granting authority can be held responsible for them. M.~B. and T.~A. acknowledge that this research was funded in whole, or in part, by the Austrian Science Fund (FWF) [P 36093]. S.~M. and M.~A.~R. acknowledge that this research was funded in whole, or in part, by the Austrian Science Fund (FWF) [P 34437]. For the purpose of open access, the author has applied a CC BY public copyright licence to any Author Accepted Manuscript version arising from this submission. D.T. received funding from the European Unions Horizon 2020 research and innovation programme under grant agreement no.\ 101004159 (SERPENTINE, \href{www.serpentine-h2020.eu}{www.serpentine-h2020.eu}). Views and opinions expressed are, however, those of the authors only and do not necessarily reflect those of the European Union or the European Research Council Executive Agency. Neither the European Union nor the granting authority can be held responsible for them. 

Solar Orbiter is a space mission run as an international collaboration between ESA and NASA and operated by ESA. Solar Orbiter magnetometer operations are funded by the UK Space Agency (grant ST/X002098/1). T.~H. is supported by STFC grant ST/W001071/1. T.N-C. thanks for the support of the Solar Orbiter and Parker Solar Probe missions, Heliophysics Guest Investigator Grant 80NSSC23K0447 and the GSFC-Heliophysics Innovation Funds. Solar Orbiter SWA data were derived from scientific sensors that were designed and created and are operated under funding provided by numerous contracts from UKSA, STFC, the Italian Space Agency, CNES, the French National Centre for Scientific Research, the Czech contribution to the ESA PRODEX programme and NASA. SO SWA work at the UCL/Mullard Space Science Laboratory is currently funded by STFC (Grant Nos. ST/W001004/1 and ST/X/002152/1). 

P.H. acknowledges the support of the PSP program office and the Office of Naval Research. Parker Solar Probe was designed, built, and is now operated by the Johns Hopkins Applied Physics Laboratory as part of NASA's Living with a Star (LWS) program (contract NNN06AA01C). Support from the LWS management and technical team has played a critical role in the success of the Parker Solar Probe mission.
The Wide-Field Imager for Parker Solar Probe (WISPR) instrument was designed, built, and is now operated by the US Naval Research Laboratory in collaboration with Johns Hopkins University/Applied Physics Laboratory, California Institute of Technology/Jet Propulsion Laboratory, University of Gottingen, Germany, Centre Spatiale de Liege, Belgium and University of Toulouse/Research Institute in Astrophysics and Planetology.  
 
\end{acknowledgments}

\clearpage
\bibliography{bibliography}{}
\bibliographystyle{aasjournal}


\end{document}